\documentclass[preprint]{aastex}

\shorttitle{Surface Alfven Wave Damping}
\shortauthors{Evans et al.}

\begin{document}


\title{Surface Alfv\'{e}n Wave Damping in a 3D Simulation of the Solar Wind}


\author{R. M. Evans\altaffilmark{1} and  M. Opher\altaffilmark{1}}
\affil{George Mason University, 4400 University Drive, MSN 3F3, Fairfax, VA 22030}

\author{V. Jatenco-Pereira\altaffilmark{2}}
\affil{Universidade de S\~{a}o Paulo, Inst Astronomico e Geofisico, Rua do Mat\~{a}o 1226, Cidade Universit\'{a}ria, BR Sao Paulo, SP 05508-900, Brazil}

\and

\author{T. I. Gombosi\altaffilmark{3}}
\affil{Center for Space Environment Modeling, University of Michigan, 2455 Hayward Street, Ann Arbor, MI 48109}


\email{revansa@gmu.edu}


\begin{abstract}
Here we investigate the contribution of surface Alfv\'{e}n wave damping to the heating of the solar wind in minima conditions. These waves are present in regions of strong inhomogeneities in density or magnetic field (e. g., the border between open and closed magnetic field lines). Using a 3-dimensional Magnetohydrodynamics (MHD) model, we calculate the surface Alfv\'{e}n wave damping contribution between 1-4 $R_{\sun}$ (solar radii), the region of interest for both acceleration and coronal heating. We consider waves with frequencies lower than those that are damped in the chromosphere and on the order of those dominating the heliosphere: $3\times 10^{-6}-10^{-1}$ Hz. In the region between open and closed field lines, within a few $R_{\sun}$ of the surface, no other major source of damping has been suggested for the low frequency waves we consider here. This work is the first to study surface Alfv\'{e}n waves in a 3D environment without assuming a priori a geometry of field lines or magnetic and density profiles. We demonstrate that projection effects from the plane of the sky to 3D are significant in the calculation of field line expansion. We determine that waves with frequencies $>$2.8 $\times 10^{-4}$ Hz are damped between 1-4 $R_{\sun}$. In quiet sun regions, surface Alfv\'{e}n waves are damped at further distances compared to active regions, thus carrying additional wave energy into the corona. We compare the surface Alfv\'{e}n wave contribution to the heating by a variable polytropic index and find that it an order of magnitude larger than needed for quiet sun regions. For active regions the contribution to the heating is tweny percent. As it has been argued that a variable gamma acts as turbulence, our results indicate that surface Alfv\'{e}n wave damping is comparable to turbulence in the lower corona. This damping mechanism should be included self consistently as an energy driver for the wind in global MHD models.  

\end{abstract}


\keywords{Sun: corona, Sun: magnetic fields, Sun: solar wind, waves}


\section{Introduction}

The physical mechanisms behind the heating of the solar corona and the acceleration of the fast solar wind are two major unresolved issues in solar physics. Thermal heating alone is not sufficient to bring models into agreement with observations of the lower corona and at Earth \citep{usm03}. The acceleration of the solar wind occurs predominantly within a few solar radii of the surface \citep{har80, gra96}. Additionally, Solar and Heliospheric Observatory (SOHO) observations have shown that ions are heated below 4 $R_{\sun}$ \citep{koh98, ess99}. \citet{gra96} suggested that because the locations for the heating of the corona and the acceleration of the solar wind are the same, it is possible that the same mechanism could contribute to both. 

One source of heating is magnetic reconnection associated with flares and nanoflares at the solar surface. Using extreme ultraviolet observations, \citet{pat09} have suggested the heating of active regions is impulsive, and therefore could be associated with nanoflares. In the high speed solar wind, reconnection events current sheets and filaments have been suggested as a heating mechanism \citep{mat03}. 

In the lower corona, the region of interest in this work, damping of Alfv\'{e}n waves is known to produce nonthermal acceleration and to bring models into agreement with observations both near the Sun and at large distances \citep{par65}. The possible damping mechanisms for Alfv\'{e}n waves in the photosphere to lower corona are numerous and include nonlinear damping \citep{wen89,ofm97}, turbulent cascade \citep{hol86,mat99}, phase mixing \citep{hey83,par91}, Landau damping \citep{hol71}, neutral collisional \citep{dep01,lea05}, ion-cyclotron damping \citep{ise01}, and surface Alfv\'{e}n wave damping \citep{ion78} (and references herein).

 Global magnetohydrodynamics (MHD) models that do not include waves employ different methods such as empirical heating functions or varied polytropic index distribution to drive and heat the solar wind and are able to match well with Ulysses, Yohkoh, Helios and Advanced Composition Explorer data \citep{mik99, gro00, rou03}.  To include Alfv\'{e}n wave damping in a realistic model of the solar environment, physical damping mechanisms must be specified. Some MHD models use Alfv\'{e}n waves without specifying a damping mechanism. Global \citep{usm06} and local \citep{cra07} models include additional equations for the waves and prescribe an empirical damping length. These studies are benchmarked with Ulysses and Helios observations. Alternatively, one can include the wave energy and momentum without damping \citep{lio09}, and match SOHO Extreme Ultraviolet and Yohkoh soft X-ray observations.

Of the numerous possible mechanisms for damping Alfv\'{e}n waves, those which we expect to be important for low frequency waves are nonlinear turbulent damping, phase mixing, surface Alfv\'{e}n waves. In this paper, we study the contribution of surface Alfv\'{e}n wave damping by utilizing a 3D global MHD simulation of a thermally-driven solar wind \citep{coh07}. In this model, the lower corona is the inner boundary (chromosphere and transition region are not resolved.) Waves are not explicitly included in the model, nor can low frequency waves be resolved due to spatial and temporal resolution limits (to be discussed further in Section 2.2). 

We consider waves with frequencies lower than those that are completely damped in the chromosphere and on the order of those dominating the heliosphere: $3\times 10^{-6}-10^{-1}$ Hz (periods 3 seconds to 3 days). Alfv\'{e}n waves have been detected in the lower layers of the solar atmosphere using both ground based observations \citep{jes09} and the Hinode spacecraft \citep{oka07, cir07, dep07} with periods 2-4 minutes ($4.2-8.3\times10^{-3}$ Hz). Ground based observations indicate the presence of  Alfv\'{e}n waves in the corona with periods of five minutes ($3.3\times10^{-3}$ Hz) \citep{tom07, tom09}. At 1 AU, the dominant wave power is in waves with periods of 1-3 hours ($9.2\times10^{-5}$ to $2.8\times10^{-4}$ Hz) \citep{bel71}. 

In this study we are interested in a damping mechanism that acts in the lower corona \citep{gra96}. In the chromosphere, Alfv\'{e}n waves with frequencies above 0.6 Hz are damped by ion-neutral collisional damping and frequencies below $10^{-2}$ Hz were unaffected \citep{dep01, lea05}. \citet{cra05} found that waves below $10^{-2}$ Hz were not damped by any mechanism in the chromosphere. \citet{cra05} also found that nonlinear damping occurred over the extended corona. \citet{ver07} found waves $10^{-6}-10^{-4}$ Hz were reflected by a gradient in the background Alfv\'{e}n profile, and dissipated  not in the lower layers of the atmosphere, but over distance of a few solar radii. In the corona, the ion-cyclotron frequency is $10^{4-6}$ Hz, so cyclotron resonance damping is not relevant for low frequency waves. Phase mixing of outgoing and reflected incoming Alfv\'{e}n waves has also been studied \citet{suz05, suz06}. 

Using a combination of three damping mechanisms (nonlinear damping, surface Alfv\'{e}n wave damping, and phase mixing), \citet{jat89} were able to match observations of mass loss rates and terminal velocities for cool, giant stars. They applied their model to the Sun and were able to obtain coronal heating and match wind velocity and Alfv\'{e}n wave power density observations in a 1D simulation \citep{jat94}. In this paper, we extend their work on surface Alfv\'{e}n wave damping in a 3D simulation of the solar corona. 

Surface Alfv\'{e}n waves form on a magnetic interface -  a finite thickness boundary separating two regions of plasma with a strong inhomogeneity in magnetic field and/or density. 
The Alfv\'{e}n wave in each region can interact, damp and transfer energy into the resonant layer separating the two plasmas (resonant absorption).  
\citet{ion78} first utilized surface Alfv\'{e}n waves and resonant absorption as a mechanism to heat coronal loops. The transfer of MHD wave energy by resonant absorption was also studied in \citet{hol87} and \citet{wen79}. An alternative dissipation mechanism for surface Alfv\'{e}n waves is nonlinear wave steepening \citep{rud92}.
These and other efforts, e.g. \citep{rob86} have resulted in damping lengths which depend on the frequency of the waves, the nature of the magnetic interface, and the local plasma parameters (density, magnetic field and velocity). 

Utilizing these relations, the profile of the damping length in the wind has been estimated \citep{jat89,nar98}. All previous studies made assumptions about the wind. For example, \citep{nar98} calculated nonlinear viscous damping of surface Alfv\'{e}n waves in polar coronal holes. They assumed two values of the superradial expansion of the magnetic field lines, profiles for density (based on observations), and a single frequency (0.01 Hz). They obtained one profile, and concluded that the nonlinear damping of the surface Alfv\'{e}n waves in region of strong magnetic field expansion should contribute significantly to the heating in the solar wind. 

The surface Alfv\'{e}n wave damping length depends on the profile of the background Alfv\'{e}n speed. As was shown in a survey of Alfv\'{e}n speed profiles from several MHD models \citep{eva08}, the Alfv\'{e}n profiles for \citet{ver07} and \citet{cra07} are almost identical below 10 $R_{\sun}$, and the profiles were different from MHD models using empirical heating functions. 
The profile for \citet{usm06} is similar to these two, but differs very low in the corona. \citet{eva08} concluded that the inclusion of Alfv\'{e}n waves with empirical damping brought MHD models better in alignment with local heating studies that had the best agreement with observations. 

In the present study, we quantify the surface Alfv\'{e}n wave damping length for use in an MHD wave-driven model.  We expect to find surface Alfv\'{e}n waves in the border between the fast and slow solar wind for two reasons: a) the gradient in density and b) the superradial expansion of the open magnetic field required to fill the space over closed streamers. We will focus on the superradial expansion of the field lines and compare our 3D MHD model with the observational study of \citet{dob99}. Our calculations show that waves with periods less than 1 hour (frequency greater than 2.8 $\times 10^{-4}$ Hz) are damped in the region between 1-4 $R_{\sun}$, the region of interest for both solar wind acceleration and coronal heating \citep{gra96}. No other major source of damping has been suggested for these waves in this region. We demonstrate the importance of the 3D geometry in our results. We show that the contribution of the damping of surface Alfv\'{e}n waves to the wind is on the order of magnitude as (or in some cases larger than) turbulence. It is important to note that this study of waves in a solar wind solution was not self consistent - we did not consider any back effects on the waves from the plasma.

The paper is organized as follows: in section 2 we describe the theory and numerical simulation background. In section 3 we calculate the location of coronal hole boundaries, and the damping length profile along those lines. In section 4 we estimate the energy surface Alfv\'{e}n waves will deposit below 4 $R_{\sun}$. We also calculate the heating invoked in a semi-empirical thermodynamical model, compare with the wave flux and discuss the results. Finally, conclusions can be found in section 5. 

\section{Methods and Data}

\subsection{Theory}
The inhomogeneity in density and/or magnetic field that gives rise to the surface Alfv\'{e}n waves can be described analytically as either a discontinuity or finite layer \citep{has82}, such as a flux tube. 
For the case of a rapidly expanding flux tube of width $a$, (where $a$ is much smaller than the radius of the flux tube), surface Alfv\'{e}n waves form on the inner and outer surfaces. These waves can interact, damp and deposit energy into the surrounding plasma with a damping rate \citep{rob86},
\begin{equation}
\label{eq:rate1}
\Gamma_{SW} = \pi(\bar{k}a) \left(\frac{\omega_{2}^2-\omega_{1}^2} {8 \omega}\right)
\end{equation}
where $\bar{k}$ is the average wave number, $\omega$ is the frequency  and $\omega_{1}$ and $\omega_{2}$ are the Alfv\'{e}n wave frequency on either side of the flux tube (1 representing inside and 2 outside.) We assume the width of the flux tube to be much smaller than the radius. This allows us to take $\bar{k}a=0.1$, as in \citet{jat89}. If the frequency on the outside is much larger than the frequency inside (i.e., a strong inhomogeneity), then the damping rate is
\begin{equation}
\label{eq:rate2}
\Gamma_{SW} = \frac{\pi\omega(\bar{k}a)}{4 \sqrt{2}}.
\end{equation}

The surface Alfv\'{e}n wave damping length can be written as the Alfv\'{e}n speed $v_{A}=\sqrt\frac{B^2}{4\pi\rho}$ divided by the damping rate,
\begin{equation}
\label{eq:L1}
L_{SW} = \frac{v_{A}}{\Gamma_{SW}} = \frac{v_{A}4 \sqrt{2}}{\omega\pi(\bar{k}a)}.
\end{equation}
The initial damping length $L_0$ can be written as
\begin{equation}
\label{eq:L01}
L_{0} = \frac{v_{A0}4 \sqrt{2}}{\omega\pi(\bar{k}a)},
\end{equation}
which, by taking $\bar{k}a=0.1$, can be simplified to
\begin{equation}
\label{eq:L0}
L_{0} = 18\frac{v_{A0}}{\omega}.
\end{equation}

Utilizing the relation that $a \propto A(r)^\frac{1}{2} \propto r^\frac{S}{2}$ (where $S$ is the superradial expansion factor of the field line), and fixing the frequency of the waves to be constant with height, the damping length in the inertial frame is 
\begin{equation}
\label{eq:L}
L_{SW} = L_{0}\left(\frac{r_{0}}{r}\right)^\frac{S}{2} \left(\frac{v_{A}}{v_{A0}}\right)^2 \left({1+M_{A}}\right)
\end{equation}
where $M_{A}=\frac{u_{SW}}{v_{A}}$ is Alfv\'{e}n Mach Number, $u_{SW}$ is the solar wind speed,  $B$ is the magnetic field strength, $\rho$ is the mass density. The subscript 0 indicates the variable is to be evaluated at the reference height. 

In the present study the model does not treat the chromosphere or the transition region; the lower corona is the inner boundary. The cell size at the inner boundary is $\frac{3}{128} R_{\sun}$, and so we chose $r_0$=1.04$R_{\sun}$ as our reference height to assure that calculations would not include the solar surface inner boundary cells. 

We quantify the expansion of open field lines by $S$, given by. 
\begin{equation}
\label{eq:S}
A_{cs}(r) = A_{cs}(r_0)\left(\frac{r}{r_0}\right)^{S(r)}
\end{equation}
where $A_{cs}(r)$ is the cross sectional area of the flux tube at distance $r$. A value of 2 for $S$ indicates pure radial expansion. The lines which border closed field lines must open faster than radial to fill the space above the closed loops. In studies where $S$ is not a function of $r$, typical values chosen were 2-6 for open field regions from 1-10 $R_{\sun}$ \citep{moo91, jat94, nar98}. Here we determine $S$ explicitly as a function of height from 1.04-10 $R_{\sun}$.

A similar parameter for the expansion of a field line is the superradial diverging factor or superradial enhancement factor $f$, as in
\begin{equation}
\label{eq:f}
A_{cs}(r) = A_{cs}(r_0)\left(\frac{r}{r_0}\right)^{2}f(r).
\end{equation}
We will use a 3D MHD model as a laboratory to estimate the contribution of the surface waves to the wind from 1.04-10 $R_{\sun}$. 
\subsection{Generation of Steady State}

We obtain steady state solar wind solutions by using the Solar Corona component of the Space Weather Modeling Framework (SWMF), developed by the University of Michigan \citep{tot05}. This 3D global magnetohydrodynamics (MHD) model incorporates Michelson Doppler Imager (MDI) magnetograms to generate an initial magnetic field configuration with the Potential Field Source Surface model (see \citet{coh07} for details). An initial density is assumed on the solar surface ($3.4 \times 10^{8}$ $ cm^{-3}$), and the MHD equations are evolved in local time steps (12,000 iterations) and time accurate calculations (for ten minutes) to achieve steady state solutions for solar minima conditions in a 24 $\times$ 24 $\times$ 24 $R_{\sun}$ domain. 

The steady state was generated with Carrington Rotation (CR) 1912. Solar wind solutions from SWMF were validated in \citet{coh08} from CR1916-1929 by comparing with Advanced Composition Explorer and Wind satellite data (near 1 AU). CR1912 was chosen to allow for comparisons to an observational study of the expansion of open field lines \citep{dob99}.

Waves occur naturally as a perturbation to the MHD equations, and so their presence may be expected when solving the MHD equations in space and time. However, in global simulations waves have to be included explicitly \citep{usm03} due to time and spatial limitations. The time step of this simulation (0.2 seconds) is less than the smallest period considered in this analysis (3 seconds). Additionally, the grid resolution is not enough to spatially resolve the waves.  

\section{Coronal Hole Boundary Analysis}

\subsection{Location and Expansion Factor}
\citet{dob99} (herein referred to as DO99) characterized the large-scale solar magnetic topology during solar minima conditions (August 1996) using data from the Solar and Heliospheric Observatory's (SOHO) Ultraviolet Coronagraph Spectrometer instrument. They analyzed the latitudinal dependence of two line emission intensities and found the values were constant within the large polar coronal holes but suddenly increased at the border of the holes and equatorial streamers. 
DO99 used this increase in intensity to identify the colatitude of coronal hole boundary (CHB), i.e.  the border between open and closed magnetic field lines. In the present study, we identify the CHB locations as the open field line with the largest colatitude (the angle measured from the pole to the equator) in the steady state described in the previous section. We compare our calculation of the field expansion with the observed values from DO99.

Figure 1 shows the 3D coronal hole boundary (CHB) field lines obtained from the model. The different colors refer to: red as northeast (NE); blue as southeast (SE); green as southwest (SW); and purple as northwest (NW). The solar surface is shown colored by the radial component of the magnetic field. On 1996 August 17, there was an active region near the center of the disk, and mostly quiet sun at the intersection of the plane of the sky with the photosphere.

The first two rows of Table \ref{tbl1} provide CHB colatitudes (angle measured from the pole to the CHB footpoint) found in DO99 and this study. We find that 3 of the 4 simulated CHB are higher in latitude (i.e., smaller colatitude) compared to DO99. As we discuss below, these differences could be due to projection effects. 

The superradial expansion factor $S$ and superradial enhancement factor $f$ were calculated as a function of height for each CHB field line. In Figure 2 we show $f$ for the CHB lines in this study. Additionally, we have included the $f$ profiles which correspond to the minimum and maximum asymptotic values of $f$ for CHBs determined in DO99. Rows 3 and 4 in Table \ref{tbl1} provide the asymptotic superradial enhancement factor value for each line. We find that both $S$ and $f$ cover a larger range of values compared to DOB99. Only the SW line from our simulation falls in the range from DO99 and only from 3-6 $R_{\sun}$. In general, we find that 3 of the 4 boundary lines (all except SW) have lower values in our simulation compared to DO99.

The best agreement for the location of the CHB between the studies is the SW line. The southern hemisphere from the simulation is more similar to the results deduced from observations in the plane of the sky in DO99 than the northern hemisphere. Overall, we find that \citet{dob99} had larger values of $f$, and a small range of values, for the same field lines. As we show below, this result is due to 3D vs. 2D projection effects. 

\subsection{Projection Effects}
In order to quantify the significance of 2D projection effects in the context of the expansion of field lines, we calculated $f$ for four field lines shown in Figure 3a. The projection of the field lines on the plane of the sky is shown in Figure 3b. Contours of solar wind speed (in $\slantfrac{km}{s})$ are shown in Figures 3a and b, and the solar surface is shown as white sphere. Figure 3c provides the superradial enhancement factor $f$, calculated according to Equation \ref{eq:f} for each field line (labeled A, B, C and D as in Figure 3b). The calculation of the 3D line is shown as solid lines, and the 2D projection line calculation is shown as dashed lines. 

Figure 3c demonstrates that a 2D projection of a 3D field line on the plane of the sky can overestimate the divergence of the line. The 3D lines from A-D all approach values of f~4 while the 2D estimates vary between 4 and 13. This projection effect explains why the values for $f$ from our simulation are smaller than those determined in the observational study of DO99. It is crucial, therefore, to do studies of surface Alfv\'{e}n wave damping in a 3D simulation in order to capture the true divergence of the field lines.

\subsection{Damping Length}
In Figure 4, we plot $L_{SW}$ (see Eq. \ref{eq:L}) which was calculated using parameters $\rho$, $B$ and $u_{sw}$ from the steady state solution. Figure 4a presents  $L_{SW}$ for the coronal hole boundary field lines in Figure 1 with frequency $4.17 \times 10^{-3}$ Hz, normalized to the initial damping length $L_{0}$ of the SW line (chosen because of the agreement with DO99). This normalization allows for comparison of the profile features from different source regions as a function of height. In Figure 4b we feature only the SW line and present $L_{SW}$ for several frequencies, from $3.3 \times 10^{-1}- 3.8 \times 10^{-6}$ Hz. It can be seen in Figure 4b that frequencies above $2.8 \times 10^{-4}$ (short dashed line) will be appreciably damped within a few solar radii of the surface.

Figure 4a shows distinctly different profiles from the southern and northern CHB lines. We examined the source region of each footpoint and found that the SE and SW lines originated near small active regions in which the radial component of the magnetic field $B_r\approx50$ G. Both northern hemisphere lines originated from quiet sun regions ($B_r\approx1$ G). For SWMF and other MHD models, \citet{eva08} showed that the Alfv\'{e}n profile will contain a maximum, or hump, if the source region is quiet sun. The profile from an active region in global models begins at a maximum value, and drops to less than a few hundred $\slantfrac{km}{s}$ within one solar radius from the surface. 

The profile of $L_{SW}$ is controlled by the Alfv\'{e}n speed profile. The normalized profiles in Fig. 4a show that the position corresponding to $L_{SW}=1R_{\sun}$ is closest to the Sun for active regions. The profiles from quiet Sun source regions have a plateau, pushing $L_{SW}=1R_{\sun}$ further from the Sun. The implication of this result can be seen in the equation relating the Alfv\'{e}n wave energy density, 
\begin{equation}
\label{eq:epsilon}
\epsilon_{SW} = \left(\frac{M_{A0}}{M_{A}}\right) \left(\frac{1+M_{A0}}{1+M_{A}}\right)^2 \exp\left(-\frac{r}{L_{SW}}\right).
\end{equation}
If the damping length is $1R_{\sun}$ or less, then the waves will be damped close to the Sun. Therefore, the presence of the hump means the energy of the surface Alfv\'{e}n wave can travel further into the corona before substantial damping occurs. This means that the quiet sun region will damp more surface waves at further distances, so it is more efficient in carrying the wave momentum out into the corona. Active regions will damp closest to the Sun. 

\section{Dissipation of Wave Energy and Heating}

\subsection{Wave Energy}
In the previous section we considered how surface Alfv\'{e}n waves at the solar surface with the frequency range $3.8\times10^{-6}$ to $3.3\times 10^{-1}$ Hz \citep{cra05} would be damped in our background solar wind environment. We found that waves with frequency above $2.8\times 10^{-4}$ Hz were appreciably damped below 4$R_{\sun}$. We now consider the contribution of their wave flux to the energy of the wind. There will be a contribution to the momentum of the wind as well, but in this analysis we ignore this contribution. We assume that the wave damping will contribute solely to heating the wind; therefore we derive here an upper limit on their contribution to the heating of the plasma.
The spectra of the surface Alfv\'{e}n waves \citep{jat94} is: 
\begin{equation}
\label{eq:WaveFlux}
\phi_{AW}(\omega) = \phi_{0} \left(\frac{\omega}{\bar{\omega}}\right)^{-\alpha} \frac{erg}{ cm^{2} s Hz}
\end{equation}
where $\phi_{0} =1.3\times 10^5 \slantfrac{erg}{cm^{2}sHz}$, $\bar{\omega}$ is the mean frequency in the observed range and the power index corresponding to the low frequency waves we are considering is $\alpha$=0.6 \citep{tu89}.

We assume that this flux of Surface Alfv\'{e}n waves is propagating along open field lines during solar minima. This flux will decrease with distance as
\begin{equation}
\label{eq:WaveFluxwithDistance}
\phi(\omega,r) = \phi_{0}\left(\frac{\omega}{\bar{\omega}}\right)^{-\alpha}\exp\left(\int{-\frac{dr}{L_{SW}(\omega,r)}}\right)
\end{equation}
where $L_{SW}$ is the damping length, for each frequency. With the damping lengths calculated in the previous section, we calculate how much flux is lost between $R=1.04$ (our self-imposed reference height) and $4R_{\sun}$. The height of $4R_{\sun}$ was chosen because the contribution to solar wind acceleration and coronal heating must be deposited within a few solar radii of the Sun \citep{koh98, ess99}.

The contribution to the solar wind at each frequency is
\begin{equation}
\label{eq:fluxlost2}
\phi_{lost}(\omega)= \phi_{0}\left(\frac{\omega}{\bar{\omega}}\right)^{-\alpha}\left(1-\exp\left(\int_{r_1}^{r_2}{\frac{-dr}{L_{0}(\omega)\left(\frac{r_{0}}{r}\right)^\frac{S}{2} \left(\frac{v_{A}}{v_{A0}}\right)^2 \left(1+M_{A}\right)}}\right)\right)
\end{equation}
where the limits are $r_1=1.04R_{\sun}$ and $r_2=4R_{\sun}$, and the definition of $L_{sw}$ from Equation \ref{eq:L} has been included ($v_{A}$, $S$, $M_{A}$, are all functions of r.)

We replace $L_{0}$ in Eq. \ref{eq:fluxlost2} with Eq. \ref{eq:L0}. The total flux lost is found using 
\begin{equation}
\label{eq:fluxlost4}
\phi_{lost,total}= \int_{\omega_1}^{\omega_2}{\phi_{0}\left(\frac{\omega}{\bar{\omega}}\right)^{-\alpha}\left[1-\exp\left(\int_{r_1}^{r_2}{-\omega\frac{r^\frac{S}{2}v_{A0}}{18 r_{0}^\frac{S}{2} v_{A}^2 \left(1+M_{A}\right)}dr}\right)\right]d\omega}
\end{equation}
where the limits are $\omega_1=2.8\times 10^{-4}$ Hz and $\omega_2=0.3$ Hz.

Next we compare the potential contribution of the wave flux to the heating in the model (Eq. \ref{eq:heat2}). It should be stressed that we are not doing a self-consistent calculation: we are estimating wave flux from a model that does not include waves, and we are not considering any feedback of the waves on the plasma. 

\subsection{Heating of the Corona}
The simulation analyzed in this paper is that of a thermally driven solar wind. The polytropic index $\Gamma$ in the model varies in space by utilizing the Wang-Sheeley-Arge model and the conservation of energy along a solar wind field line using the Bernoulli equation. This serves to artificially heat the wind \citep{coh07} in a manner mimicking turbulence \citet{rou03}. Figure 5 shows the distribution of $\Gamma$ in the plane of the sky on 1996 August 17. In this section we quantify the heating due to this variable gamma, and do a comparison with the energy deposited by damped surface Alfv\'{e}n waves, as in Eq. \ref{eq:fluxlost4}.

The first law of thermodynamics attributes changes in the internal energy $U$ of a gas to work done on or by the gas $W$, and heat added to or removed from the gas $Q$. In the case of an ideal gas, the change in internal energy can be written as $dU=c_{v}dT$, where $c_{v}$ is the specific heat at constant volume. The work is expressed as $dW=-pdV$, where $p$ is the pressure and $V$ is the volume. The first law can therefore be written as
\begin{equation}
\label{eq:firstlaw}
dQ=c_{v}dT+pdV.
\end{equation}
By introducing the ideal gas equation of state and assuming that the ratios of specific heats are constants, one can derive a polytropic equation,
\begin{equation}
\label{eq:poly}
\frac{p}{\rho^{\frac{c_{p}-c}{c_{v}-c}}}=\frac{p}{\rho^{\alpha}}=const.
\end{equation}
where $\alpha$ is referred to as the polytropic index. The notation stems from \citet{par63} to clarify that this index can (but need not) be the ratio of specific heats, and that we are not necessarily considering an adiabatic process. The symbol $\gamma$ is typically used for the ratio of specific heats, and in the case of an adiabatic expansion (no heating enters or leaves the system), $\alpha=\frac{5}{3}$. An isothermal wind expansion would be characterized by $\alpha=1$. Observations of the solar wind have indicated that $\alpha$=1.46-1.58 in the heliosphere \citep{tot95}. A value closer to unity is adopted in some global MHD models in the region near the Sun in order to generate fast solar wind and match temperature observations in the heliosphere \citep{usm03}. 

All previous discussion had the underlying assumption that $\alpha$ was constant with height. If that condition is not met, then the polytropic index is referred to as an effective (or local) polytropic index and written as $\Gamma$ \citep{tot95}. The polytropic equation (Eq. \ref{eq:poly}) is modified to
\begin{equation}
\label{eq:modpoly}
\frac{dlnP}{dr}=\Gamma\frac{dln\rho}{dr}+ln\rho\frac{d\Gamma}{dr}
\end{equation}
such that the relationship between density and pressure is not simple.
The variation of $\Gamma$ with height has been utilized to heat the solar wind used in this paper. We will characterize the additional heating provided by the prescribed distribution of $\Gamma$ in our solar wind simulation, and argue that surface Alfv\'{e}n waves damped near the Sun could replace this artificial heating and move the model towards a more physical treatment of the solar environment. 
For a solar wind with a constant ratio of specific heats $\gamma$, the conservation of energy can be written as \citep{man04}:
\begin{equation}
\label{eq:energy_conserve}
\frac{\partial\varepsilon}{\partial t} +  \nabla \cdot \left[ {\bf u} \left( \varepsilon + p + \frac{B^2}{8\pi} \right) - \frac{\left({\bf u}\cdot{\bf B}\right) {\bf B}}{4\pi}\right] = \rho {\bf g}\cdot {\bf u} + q
\end{equation}
where p is the thermal pressure, q is the additional heating function, and the energy density is
\begin{equation}
\label{eq:enden}
\varepsilon=\frac{\rho u^2}{2}+\frac{p}{\gamma-1}+\frac{B^2}{8\pi} .
\end{equation}
Recent global MHD studies adopt an exponential function for the form of q with several free parameters in order to benchmark the model with observations during solar minima conditions \citep{gro00, man04}.
Substituting Eq. \ref{eq:enden} into Eq. \ref{eq:energy_conserve}, and setting time derivatives to zero for a steady solar wind, we find:
\begin{equation}
\label{eq:energy_conserve2}
\nabla \cdot \left[ {\bf u} \left( \frac{\gamma p}{\gamma -1} + \frac{\rho u^2}{2}+\frac{B^2}{4\pi} \right) - \frac{\left({\bf u}\cdot{\bf B}\right) {\bf B}}{4\pi}\right] = \rho {\bf g}\cdot {\bf u} + q.
\end{equation}
As discussed at the beginning of this section, there is no heating function q in the simulation used in this paper, and the ratio of specific heats $\gamma$ is replaced by the effective gamma  $\Gamma$,
\begin{equation}
\label{eq:energy_conserve3}
\nabla \cdot \left[ {\bf u} \left( \frac{\Gamma p}{\Gamma -1} + \frac{\rho u^2}{2} +\frac{B^2}{4\pi} \right) - \frac{\left({\bf u}\cdot{\bf B}\right) {\bf B}}{4\pi}\right] = \rho {\bf g}\cdot {\bf u}.
\end{equation}
Although $\Gamma$ has both latitudinal and azimuthal dependence below 4$R_{\sun}$, we consider only at the radial variation, and so we replace $\nabla$ by $\frac{d}{dr}$. We assume that we have exactly the same solar wind solution in the two cases we are considering: that of a variable gamma and of an additional volumetric heating function with $\gamma=\frac{5}{3}$. In order to quantify the amount of heating in the model with variable gamma, we subtract equation  \ref{eq:energy_conserve3} from \ref{eq:energy_conserve2}:
\begin{equation}
\label{eq:heat1}
q=-\frac{d}{dr}\left[ u_{r} \left( \frac{\Gamma p}{\Gamma -1}- \frac{\gamma p}{\gamma -1}\right) \right]. 
\end{equation}
Equation \ref{eq:heat1} can be written as:
\begin{equation}
\label{eq:heat2}
q=-\left[\frac{d\left(u_{r}p\right)}{dr}\left( \frac{\Gamma}{\Gamma -1}- \frac{\gamma}{\gamma -1}\right) - \left(\frac{d\Gamma}{dr}\frac{u_{r}p}{\left(\Gamma-1\right)^2}\right) \right].
\end{equation}

Knowing how $\Gamma$, $p$ and $u_{r}$ vary along any radial line, and setting $\gamma=\frac{5}{3}$ we can integrate equation \ref{eq:heat2} between $r_{1}$=1.04$R_{\sun}$ and $r_2$=4$R_{\sun}$ to find the heat input along any field line:
\begin{equation}
\label{eq:heat3}
Q= \int_{r_1}^{r_2}{q}dr \slantfrac{erg}{cm^2s}.
\end{equation}
This equation gives the heat deposited into the system between the two heights. We compare $Q$ with the flux of damped surface Alfv\'{e}n waves (Equation \ref{eq:fluxlost4}.) 

Table \ref{tbl1} provides $\phi_{lost}$ and $Q$ for the CHB field lines. The expansion $f$ of the NW line from the simulation has the best match to the observations (5.55 compared to 6.0). The surface Alfv\'{e}n wave flux along this line, and along the NE lin, is larger than the heating Q by an order of magnitude. The geometrical properties of the SW line also match well with observations, however the wave flux for it (and also the SE line) account for 20 $\%$ of the required heating. The distinction between the southern and northern lines is the source region: they come from a stronger magnetic field region. Near an active region, the second term in Eq. \ref{eq:heat2} (which includes the pressure and radial velocity) is larger than a quiet sun region. Therefore, we expect Q to be larger than the surface Alfv\'{e}n wave flux along lines from active regions.

As we assumed all of the wave flux goes to heating, this procedure gives an upper limit on the contribution of the damping of surface Alfv\'{e}n waves along an open magnetic field line to the heating along that line. A random sampling of 7 open field lines in the northern hemisphere with footpoints in the plane of the sky (see Figure 3) yielded $\phi_{lost}$ that were on the order, or an order of magnitude larger than the $Q$.

\section{Conclusions}
This work is the first study to look at surface Alfv\'{e}n waves in a 3D environment without assuming a priori a geometry of the field lines or magnetic and density profiles and strengths. We showed the calculation of the expansion of field lines must be done in a 3D environment. Our calculations show that waves with periods less than 1 hour (frequency greater than 2.8 $\times 10^{-4}$ Hz) are damped in the region between 1-4 $R_{\sun}$, the region of interest for both solar wind acceleration and coronal heating \citep{gra96}. We showed that the quiet sun region will damp surface waves at further distances, so it is more efficient in carrying the wave momentum out into the corona. Surface waves formed on flux tubes with footpoints in an active region will damp closer to the Sun. The required heating from an active region was found to be larger than the damping of surface Alfv\'{e}n wave flux, therefore another mechanism (such as turbulence) may be the dominant heating in these regions, with surface Alfv\'{e}n waves contributing approximately 20$\%$ of the heating.

We estimated damping of surface Alfv\'{e}n waves in the border between open and closed field lines at heights 1.04-4 $R_{\sun}$ due to the superradial expansion of the field lines. As some of the wave flux would go to the momentum of the wind, we provide an upper limit on the contribution of surface Alfv\'{e}n waves to the heating of the solar wind. In the region between open and closed field lines, within a few solar radii of the surface, no other major source of damping has been suggested for the low frequency waves we consider here. 

Our results demonstrate that it is not necessary to have turbulence in order to heat the solar wind - and that it is imperative to include the physics of surface Alfv\'{e}n wave damping in solar wind models in order to more physically model the heating. Surface Alfv\'{e}n waves could also be present in the solar wind, in the flux tube structures said to fill interplanetary space \citep{bor08}. Another environment which could support these waves are Corotating Interaction Regions (CIRs), due to the inhomogeneity in density present in these structures \citep{tsu09}. Both of these topics will be addressed in future works. 

It is important to note that this study of waves in a solar wind solution was not self consistent - we did not consider any back effects on the waves from the plasma. We simply tried to estimate if the waves could produce the heating required to create the solar wind solution from the model. In the future we will pursue other damping mechanisms with the goal of incorporating the key mechanisms of wave damping in self-consistently in global MHD models to improve the lower corona. 

We would like to thank NASA Ames for the use of the Columbia supercomputer. This research was supported by the NSF CAREER Grant ATM 0747654 and LWS NNGO6GB53G
.

\clearpage 
\begin{figure}
\includegraphics[scale=0.65]{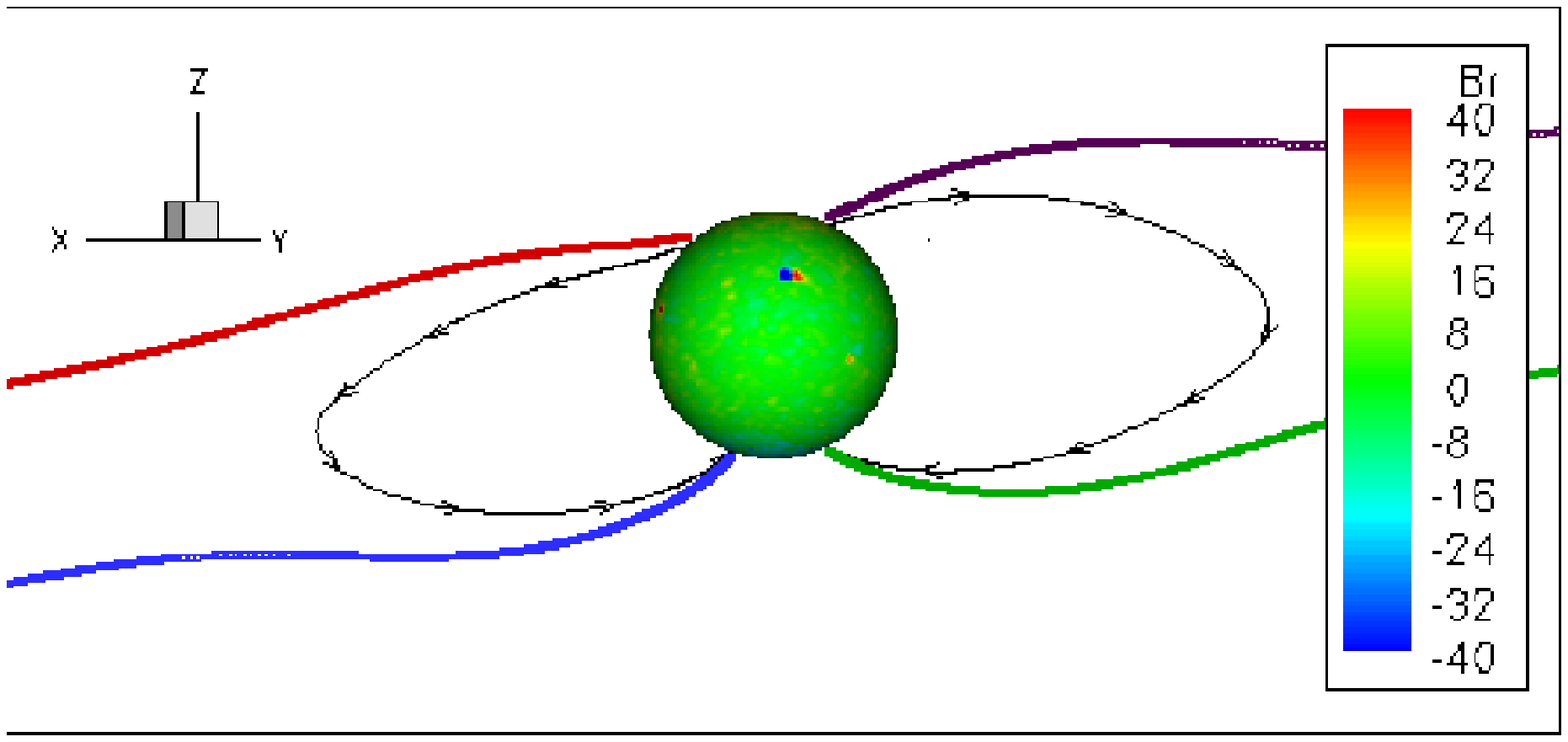}
\caption{3D coronal hole boundary field lines obtained from the simulation for 1996 August 17. The different colors refer to: red as northeast line (NE); blue as southeast line (SE); green as southwest line (SW);and purple as northwest line (NW). The black lines show closed equatorial streamers, and the arrows give the direction of the magnetic field. The solar surface is shown colored by the radial component of the magnetic field. \label{fig1}}
\end{figure}

\begin{figure}
\includegraphics[scale=0.7]{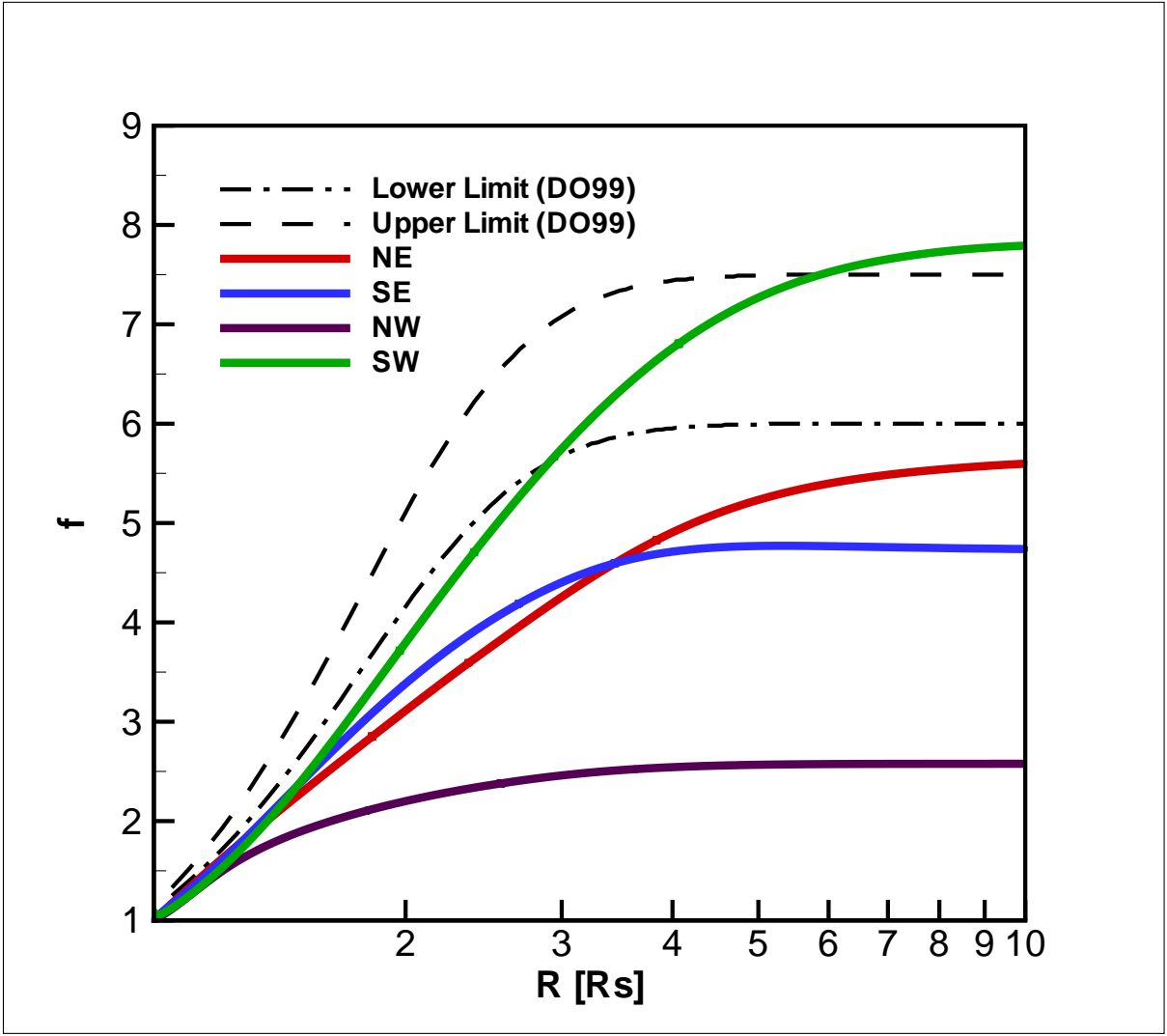}
\caption{The superradial expansion factor $f$ (see Eq. \ref{eq:f}) for the lines in Fig. \ref{fig1} (colors correspond) from 1.04-10 solar radii (Rs). We also show the profiles which correspond to the minimum (dashed line) and maximum (dash-dot line) values of $f$ determined in the observational study by \citet{dob99}. We find that $f$ covers a larger range of values compared to the observational study. \label{fig2}}
\end{figure}

\begin{figure}
\includegraphics[scale=0.3]{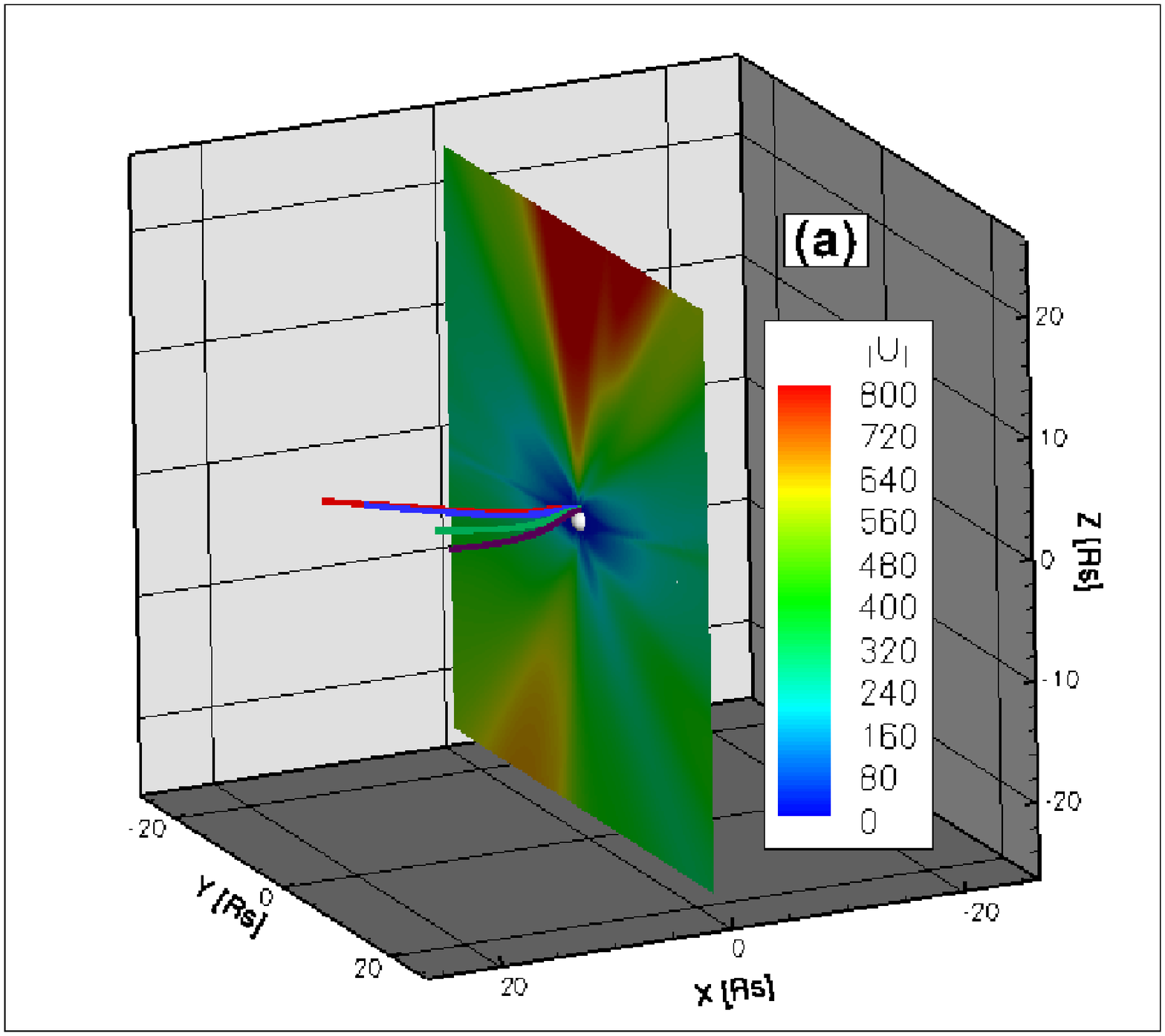}
\includegraphics[scale=0.3]{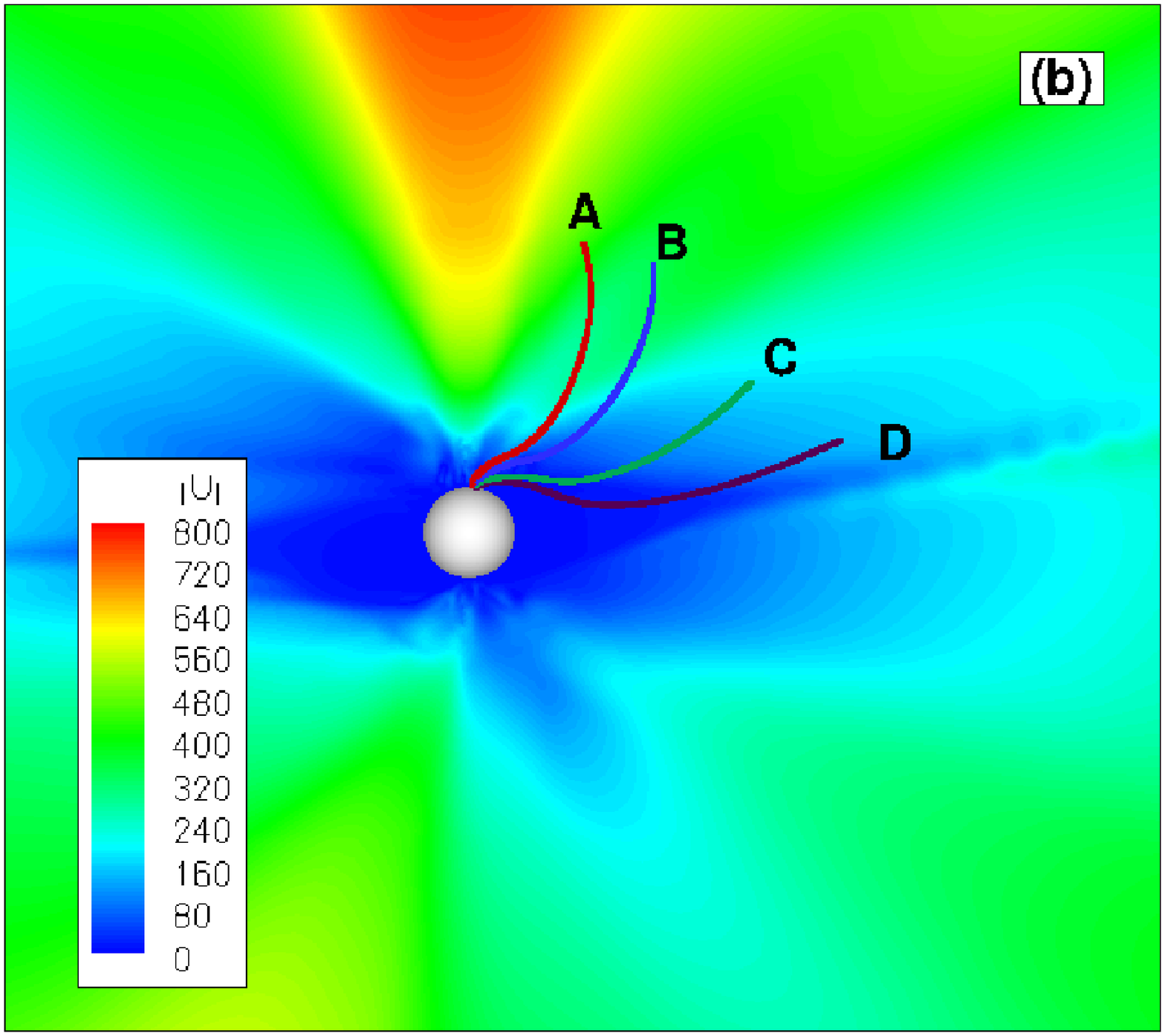}
\includegraphics[scale=0.3]{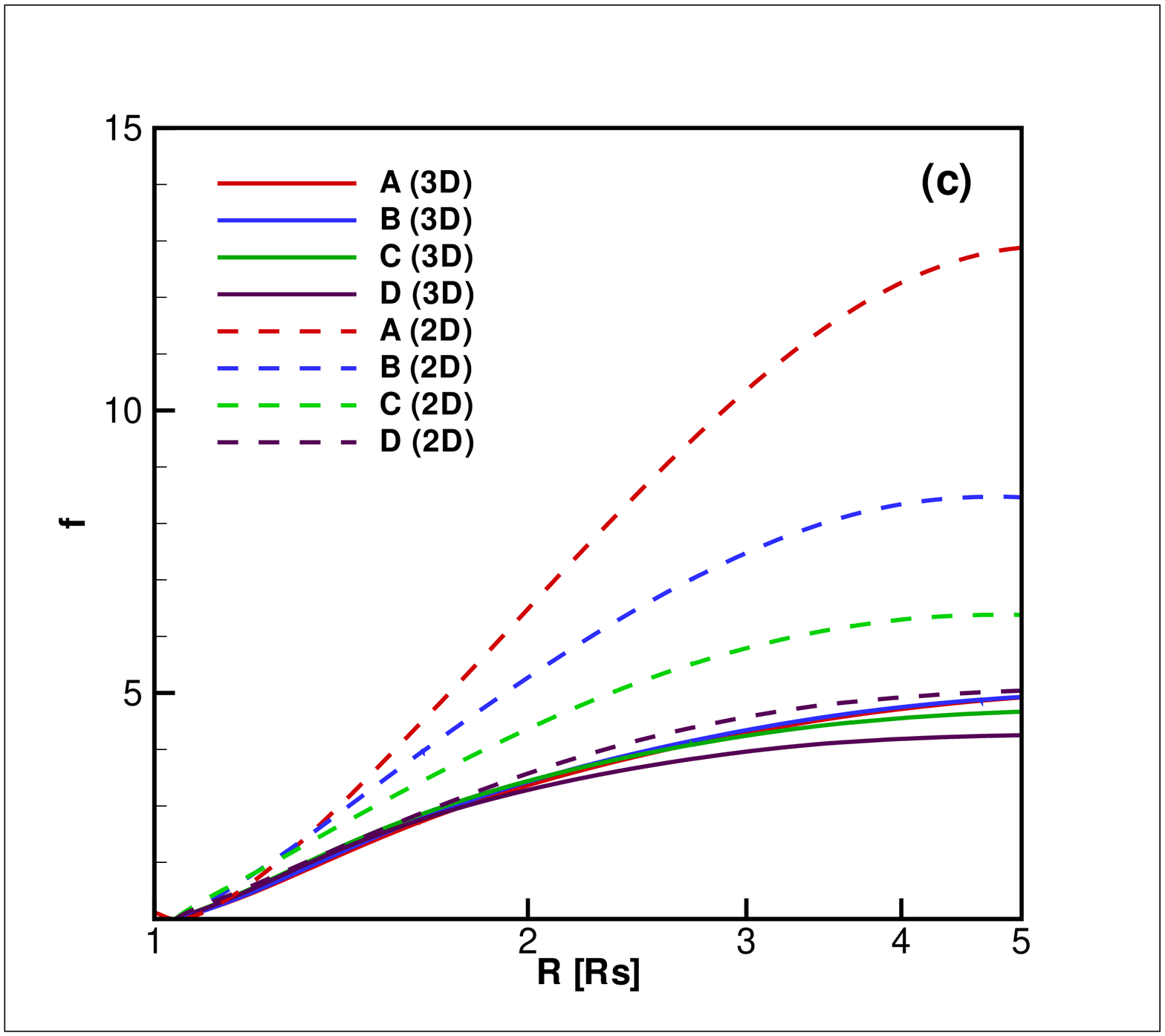}
\caption{Four field lines (A, B, C and D) shown (a) in 3D and (b) as their projections on the 2D plane of the sky. Contours of solar wind speed (in $\frac{km}{s}$) are shown on the plane, and the solar surface is shown as a white sphere. The expansion factor calculated from the 3D field line and 2D projections for each line is shown in (c). The 2D projection can overestimate the expansion factor as compared to the 3D calculation. \label{fig3}}
\end{figure}

\begin{figure}
\includegraphics[scale=0.4]{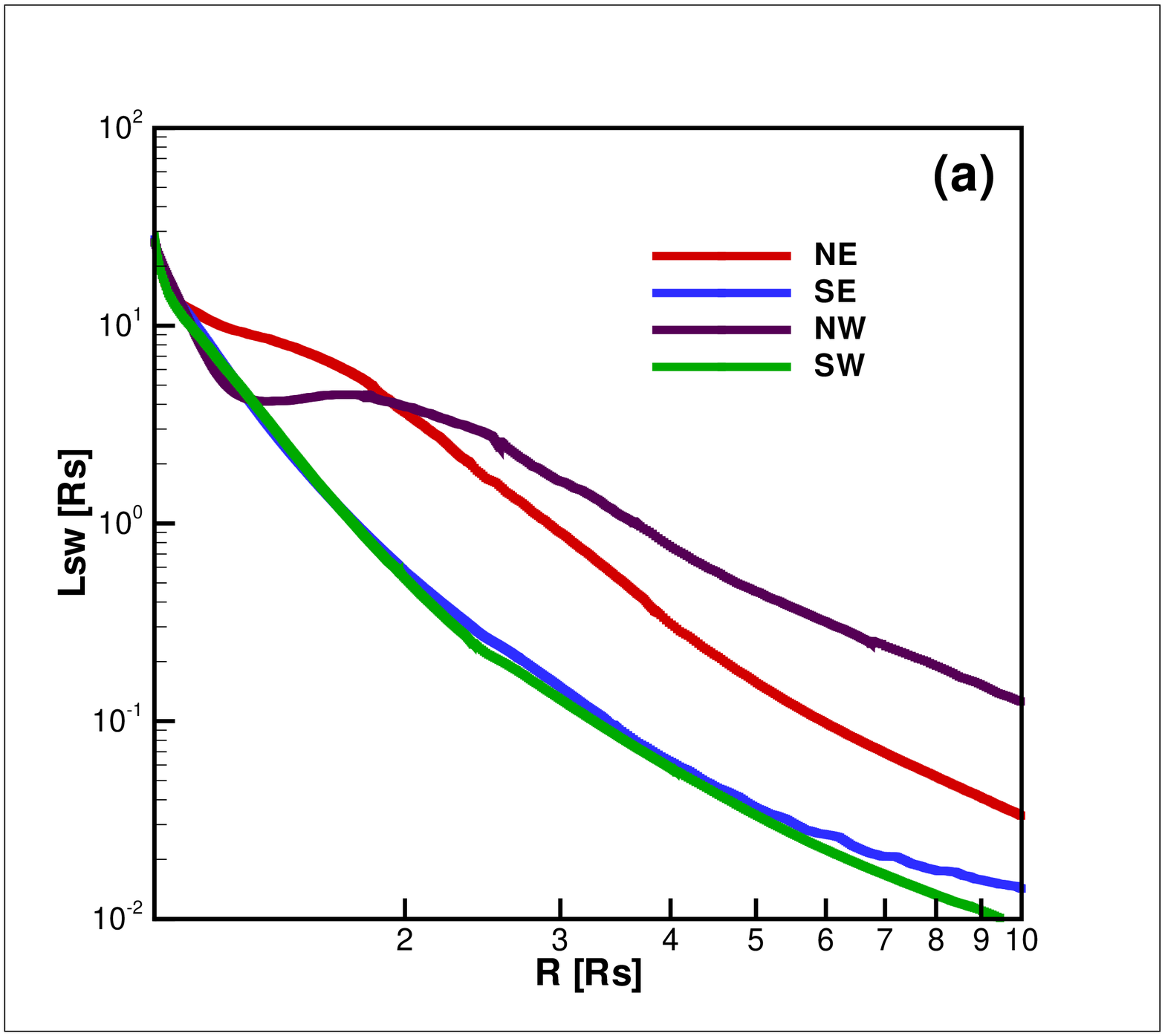}
\includegraphics[scale=0.4]{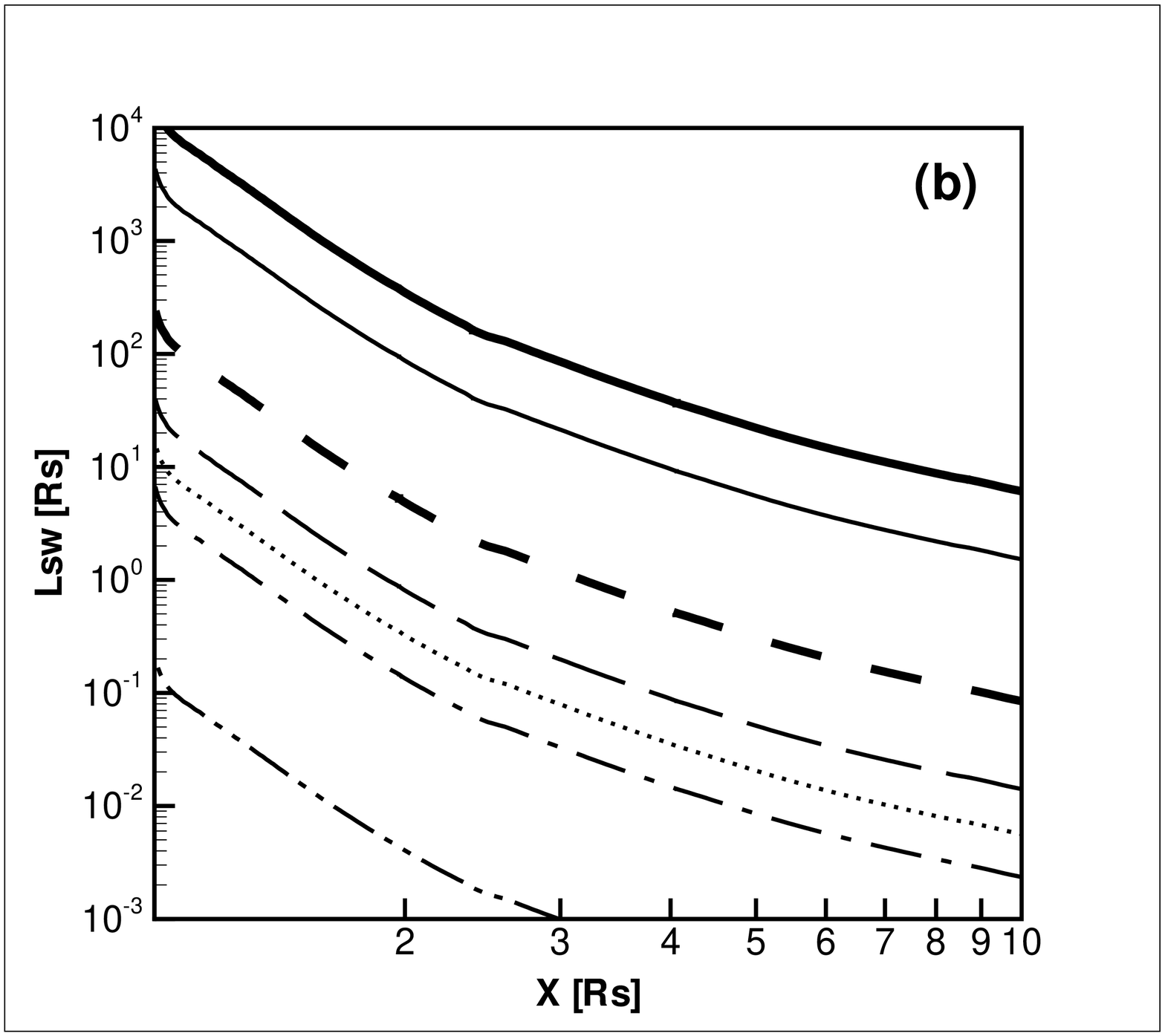}
\caption{a) Surface Alfv\'{e}n damping length ($L_{sw}$) profiles for the coronal hole boundaries in Fig. 1 (colors correspond), normalized to the $L_0$ of the SW line (green) from 1.04-10 solar radii (Rs). Note the plateau in the profiles for lines whose source region on the sun is quiet sun, differing from field lines with footpoints near active regions, whose profiles drop quickly. b) Profiles for the SW coronal hole boundary field line corresponding to different frequencies. From the bottom up to top profile: $3.3 \times 10^{-1}$ Hz (dash-dot-dot line); $1 \times 10^{-2}$ Hz (dash-dot line); $4.17 \times 10^{-3}$ Hz (dot line); $1.67 \times 10^{-4}$ Hz (long dash line); $2.8 \times 10^{-4}$ Hz (short dash line); $1.5 \times 10^{-5}$ Hz (thin solid line); $3.8 \times 10^{-6}$ Hz (thick solid line). Waves with frequencies above $2.8 \times 10^{-4}$ are appreciably damped below 4Rs. \label{fig4}}
\end{figure}

\begin{figure}
\includegraphics[scale=0.6]{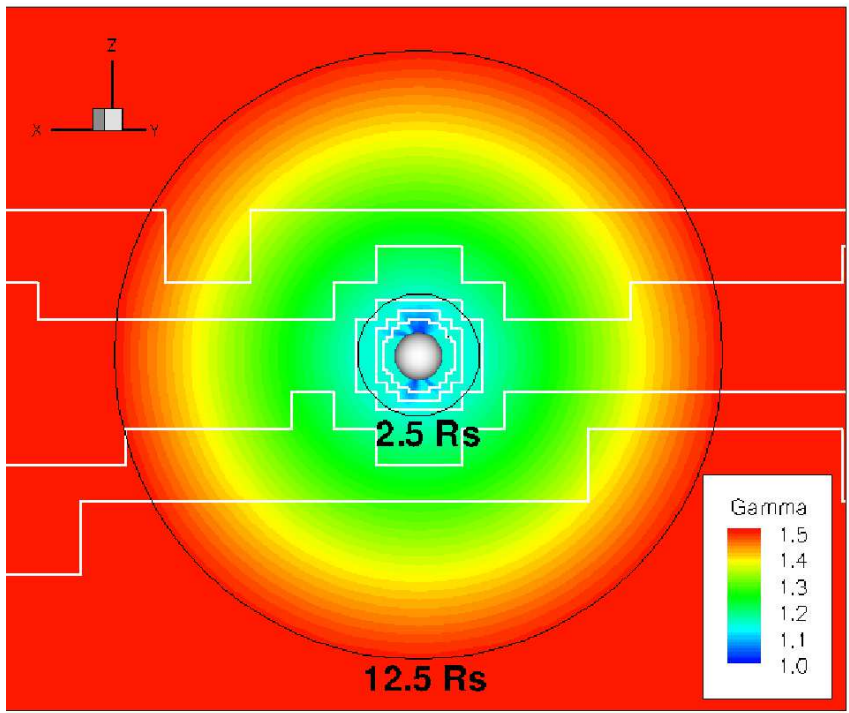}
\caption{The distribution of $\Gamma$, the effective polytropic index, in the plane of the sky on 1996 August 17. $\Gamma$ is specified on the solar surface (shown as white sphere) using the Bernoulli integral, and is interpolated to 1.1 at 2.5 Rs (inner black circle). Above 2.5Rs, $\Gamma$ varies linearly until 12.5Rs (outer black circle), above which it has the value 1.5. The white lines show the boundaries of grid refinement. \label{fig5}}
\end{figure}

\clearpage 
\begin{table}
\begin{center}
\caption{Properties of Coronal Hole Boundary Field Lines\label{tbl1}}
\begin{tabular}{crrrrrrrrrrr}
\tableline\tableline
 & NE & NW & SE & SW \\
\tableline
$\theta_{0}$\tablenotemark{a} \citep{dob99} &29.0$\degr$ &31.0$\degr$ &23.3$\degr$  &28.0$\degr$\\
$\theta_{0}$ This Study &37.2$\degr$ &22.3$\degr$ &19.2$\degr$ &25.2$\degr$\\
$f_{10R_{\sun}}$\tablenotemark{b} \citep{dob99} &6.56 &6.00  &7.30  &6.5\\
$f_{10R_{\sun}}$ This Study &2.56 &5.55 &4.71 &7.62\\
$Q\tablenotemark{c}  (\slantfrac{erg}{cm^2s})$ &$8.9 \times 10^{3}$  &$8.0 \times 10^{3}$ &$2.7 \times 10^{5}$ & $3.1 \times 10^{5}$\\
$\phi_{lost}$\tablenotemark{d} $(\slantfrac{erg}{cm^2s})$ &$6.1 \times 10^{4}$ &$6.2 \times 10^{4}$  &$6.4 \times 10^{4}$  &$6.4 \times 10^{4}$\\
\tableline
\end{tabular}
\tablenotetext{a}{Colatitude of the line - angle measured from the pole to the footpoint on the solar surface.}
\tablenotetext{b}{Value for the superradial enhancement factor at R=10 $R_{\sun}$}
\tablenotetext{c}{Heating calculated along the field line (see section 4.2)}
\tablenotetext{d}{Alfv\'{e}n wave flux deposited into the wind (see section 4.1)}

\end{center}
\end{table}

\end{document}